\documentclass[aps,pra,preprint]{revtex4}
\usepackage{graphicx}
\usepackage{latexsym}
\usepackage{bm}       
\usepackage{amssymb}
\usepackage{epsfig}
\usepackage[dvips]{color}
\usepackage{epstopdf}
\usepackage{amsmath}
\usepackage{mathrsfs}
\usepackage{appendix}
\usepackage{subfigure}
\usepackage{array}
\usepackage{multirow}

\begin{document}

\title{Response function analysis of excited-state kinetic energy functional constructed by splitting $k$-space}

\author{M. Hemanadhan and Manoj K. Harbola}
\affiliation{Department of Physics, Indian Institute of Technology, Kanpur 208 016, India}

\vskip 0.25cm

\begin{abstract}
 Over the past decade, fundamentals of time independent density functional theory for excited state have been established. 
However, construction of the corresponding energy functionals for excited states remains a challenging problem. 
We have developed a method for constructing functionals for excited states by splitting $k$-space according to the occupation of orbitals.
In this paper we first show the accuracy of kinetic energy functional thus obtained. We then perform a response function analysis of 
the kinetic energy functional  proposed by us and show why method of splitting the $k$-space could be the method of choice for construction of  energy 
functionals for excited states.
\end{abstract}

\maketitle

\section{INTRODUCTION}

Time-dependent density functional theory (TDDFT) is now a standard tool for calculating excited-state energies. Simultaneously parallel 
efforts have always been made to develop time-independent excited state density functional theory (DFT). These include work of Ziegler et al. 
\cite{ziegler}, Gunnarsson and Lundqvist \cite{gunnar}, von Barth \cite{von}, Perdew and Levy  \cite{levper}, Pathak \cite{pathak}, 
Theophilou \cite{theophi}, Oliveira, Gross and Kohn \cite{grossolkohn,olgrosskohn}, Nagy \cite{nagy}, Sen \cite{sen}, Singh and Deb \cite{debsingh}. 
Formal development \cite{levynag,gor,samal2,harbola2004} akin to ground-state density functional theory, however, is more recent. 
A challenging problem in excited state DFT is the construction of excited state energy functionals. Without such functionals, 
there is no option but to use the existing ground state energy functionals for excited states also. In such a case, the only way any 
information about the excited state can be put into energy is through the density alone \cite{von,harbola2002}. 
However, employing ground state functionals for excited states is both qualitative incorrect as well as  numerically inaccurate 
\cite{hemanadhan,hsamal,shamim}. Thus there is a need for development of appropriate excited-state energy functionals.

\begin{figure}
\begin{center}
  \includegraphics[width=5in,height=2.0in]{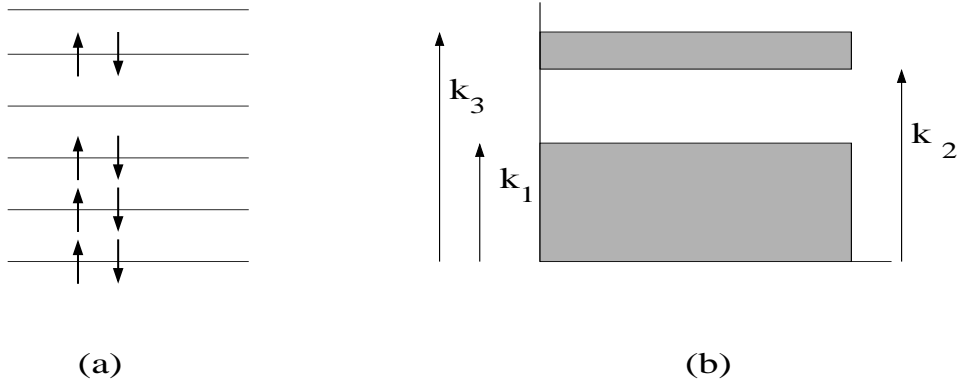}
  \caption{ Orbital occupation of electrons and the corresponding k-space occupation for an excited state
(Core-Shell) configuration similar to that of homogeneous electron gas (HEG).
}
  \label{figure}
\end{center}
\end{figure}

\begin{figure}
\begin{center}
\includegraphics[width=5in,height=2.0in]{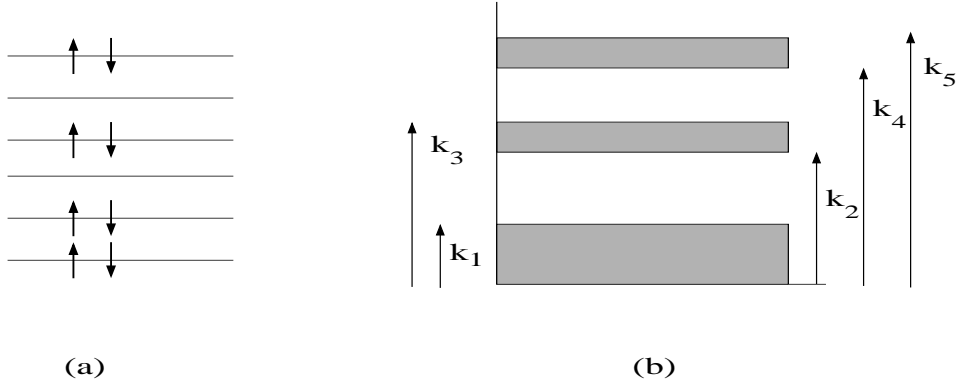}
\caption{ Excited state orbital occupation of electrons for two gap and the corresponding k-space occupation similar to that 
of homogeneous electron gas (HEG).
}
\label{core-shell-shell}
\end{center}
\end{figure}

Starting point for most of the ground state functionals has been the local-density approximation (LDA). Building on this, GEA  \cite{vonW,hermann} 
and GGA \cite{gga2,becke}  are further constructed. Thus for excited states too one would like to first develop an LDA functional and 
then improve upon it by including gradient corrections. In the past, we have proposed a systematic method of constructing  
LDA functionals for excited states. This is done by splitting the $k$-space according to the orbital occupation of a given excited state 
\cite{samal2,samalh,shamim,hsamal,aip,hemanadhan}. For example, in Figure \ref{figure} we  show  an excited state where 
some orbitals (core) are occupied, then some orbitals are vacant followed by orbitals that are occupied (shell). The corresponding 
$k$-space, also shown in Figure \ref{figure}, is constructed according to the orbital occupation such that it is occupied 
from $0$ to $k_1$, vacant from $k_1$ to $k_2$ and again occupied from $k_2$ to $k_3$  where $k_1$, $k_2$, $k_3$ are determined by 
\begin{equation}
   k_{1}^{3}(\textbf{r}) = 3\pi^{2}\rho_{c}(\textbf{r})
   \nonumber
\end{equation}
\begin{equation}
   k_{2}^{3}(\textbf{r})-k_{1}^{3}(\textbf{r}) = 3\pi^{2}\rho_{v}(\textbf{r})
   \nonumber
\end{equation}
\begin{equation}
   k_{3}^{3}(\textbf{r})-k_{2}^{3}(\textbf{r}) = 3\pi^{2}\rho_{s}(\textbf{r})
   \label{eq:k3}
\end{equation}
in terms of  $\rho_{c}$, $\rho_{v}$ and $\rho_{s}$ corresponding to the electron densities of core, vacant (unoccupied) and the shell
orbitals.  Further,
\begin{equation}
  \rho_{c}(\textbf{r}) = \sum\limits^{n_1}_{i=1} {\left| \phi_{i}^{core}(\bf{r})\right|}^{2}
  \nonumber
\end{equation}
\begin{equation}
  \rho_{v}(\textbf{r})  =  \sum\limits^{n_2}_{i=n_1+1} {\left| \phi_{i}^{unocc}(\bf{r})\right|}^{2}
  \nonumber
\end{equation}
\begin{equation}
  \rho_{s}(\textbf{r}) = \sum\limits^{n_3}_{i=n_2+1}  {\left| \phi_{i}^{shell}(\bf{r})\right|}^{2}
\end{equation}
where first $n_1$ orbitals are occupied, $n_1+1$ to $n_2$ are vacant followed by occupied orbitals from $n_2+1$ to $n_3$.
The total electron density $\rho({\bf r})$ is given as
\begin {equation}
  \rho({\bf r})=\rho_{c}({\bf r})+\rho_{s}({\bf r}) 
  \nonumber
%\label{eq:den}
\end {equation}
\begin{equation}
  \ \  \textrm{or}  \hspace{2em}  \rho({\bf r})=\rho_1({\bf r}) -\rho_2({\bf r}) + \rho_3({\bf r})
%\label{eq:den1-2-3}
\end{equation}
with $\rho_1=\rho_c,\rho_2=\rho_c+\rho_v $ and  $\rho_3=\rho_c+\rho_v+\rho_s$. 
Approximate non-interacting kinetic energy functionals for such excited-states are constructed using the ground state energy functionals.
Thus while up to second order in $\vec{\nabla} \rho $, the ground state non-interacting kinetic energy is approximated as
\begin{equation}
  \sum\limits_{i} \left \langle \phi_{i}   \right|  -\frac{1}{2}  \nabla^2  \left | \phi_{i} \right \rangle =  T^{(0)}_s[\rho] + T^{(2)}_s[\rho]
  \nonumber
\end{equation}
 where

\begin{eqnarray}
T^{(0)}_s[\rho] &=& \frac{1}{10\pi^2}\int k^5_F({\bf r})   d{\bf r}  \nonumber \\
                &=& \frac{3}{10}(3\pi^2)^{\frac{2}{3}}\int\rho^{\frac{5}{3}}({\bf r})   d{\bf r}
\label{eq:tf}
\end{eqnarray}
is the Thomas-Fermi kinetic energy functional with $k_F(\mathbf{r})=(3\pi^2 \rho(\mathbf{r}))^{1/3}$ and
\begin{equation}
T^{(2)}_s[\rho] = \frac{1}{72}\int\frac{|\nabla\rho({\bf r})|^2}{\rho({\bf r})} d{\bf r}
\label{eq:gea2}
\end{equation}
 is the second-order gradient correction. The same functional cannot be expected to be as good for excited states as it is for the ground states.
 For  excited states, the appropriate zeroth order approximation $T^{*(0)}$ is given \cite{hemanadhan} by the modified Thomas-Fermi kinetic energy functional 
\begin{equation}
T^{*(0)}[k_1,k_2,k_3]=\frac{1}{10\pi^2}\int \left(k^5_1({\bf r})-k^5_2({\bf r})+k^5_3({\bf r})
\right)d{\bf r}
\label{eq:exke0}
\end{equation}
 where the core term 
$\sum_{i=1}^{n_1}  \left \langle \phi_{i}^{core}   \right|  -\frac{1}{2}  \nabla^2  \left | \phi_{i}^{core} \right \rangle$ 
is approximated by
$\frac{1}{10\pi^2} \int k^5_1(\mathbf{r}) \ d\mathbf{r}$
and the shell part
$\sum_{i=n_2+1}^{n_3}  \left \langle \phi_{i}^{shell}   \right|  -\frac{1}{2}  \nabla^2  \left | \phi_{i}^{shell} \right \rangle$ 
is approximated by
$\frac{1}{10\pi^2} \int \left[ k^5_3(\mathbf{r})  - k^5_2(\mathbf{r})  \right] d\mathbf{r}$.
 Similarly, the construction is extended to the gradient correction up to the second-order   by writing 
\begin{equation}
 T^{*(2)} =   
  \frac{1}{72} \int    \frac{\left|\nabla \rho_1(\mathbf{r}) \right|^2   }{\rho_1(\mathbf{r})}   d\mathbf{r} 
- \frac{1}{72} \int    \frac{\left|\nabla \rho_2(\mathbf{r}) \right|^2   }{\rho_2(\mathbf{r})}   d\mathbf{r} 
+ \frac{1}{72} \int    \frac{\left|\nabla \rho_3(\mathbf{r}) \right|^2   }{\rho_3(\mathbf{r})}   d\mathbf{r} 
\label{T2*}
\end{equation}
Results obtained with these excited-state functionals for excited-states are far superior \cite{hemanadhan} to 
those calculated with the use of the ground-state functionals .
 The method is quite general and is easily extended to excited state with more than one gap.
%The method can also be applied easily to systems with two gap. 
For example, consider an excited state with two gaps with the orbital occupation of electrons
 shown in Figure \ref{core-shell-shell}. The corresponding split $k$-space is 
occupied  from $0$ to $k_1$, from $k_2$ to $k_3$ and from $k_4$ to $k_5$ 
 (core-shell-shell structure). For these excited states, the kinetic energy functional up to the second order 
 is  
\begin{equation}
T^{*(0)}(k_1,k_2,k_3,k_4,k_5)=\frac{1}{10\pi^2}\int \left(k^5_1({\bf r})-k^5_2({\bf r})+k^5_3({\bf r})-k^5_4({\bf r})+k^5_5({\bf r})
\right)d{\bf r}
\label{eq:exke0-2gap}
\end{equation}
\begin{equation}
 T^{*(2)} =   
  \frac{1}{72} \int   
  \left\lbrace
    \frac{\left|\nabla \rho_1(\mathbf{r}) \right|^2   }{\rho_1(\mathbf{r})}  
  - \frac{\left|\nabla \rho_2(\mathbf{r}) \right|^2   }{\rho_2(\mathbf{r})}  
  + \frac{\left|\nabla \rho_3(\mathbf{r}) \right|^2   }{\rho_3(\mathbf{r})}  
  - \frac{\left|\nabla \rho_4(\mathbf{r}) \right|^2   }{\rho_4(\mathbf{r})}  
  + \frac{\left|\nabla \rho_5(\mathbf{r}) \right|^2   }{\rho_5(\mathbf{r})}  
  \right\rbrace d\mathbf{r} 
\label{T2*-2gap}
\end{equation}
These functionals for spin-compensated systems (unpolarized) are easily generalized to their spin-density counterparts by writing
\begin{equation}
T^*[\rho_\uparrow,\rho_\downarrow] = \frac{1}{2}(T^{*}_\textrm{unpol}[2\rho_\uparrow]+T^{*}_\textrm{unpol}[2\rho_\downarrow])
\label{eq:lsd}
\end{equation}
where $\rho_\uparrow$ and $\rho_\downarrow$ denote the density of the up and the down spin electrons, respectively.
Results obtained with these functionals for two-gap excited states are shown in Table \ref{tab-co-sh-sh}, where we compare the excited kinetic energies calculated using \eqref{eq:exke0-2gap} and \eqref{T2*-2gap} 
 with the exact Kohn-Sham energies obtained by solving the Kohn-Sham equation with 
Gunnarsson-Lundquist parametrization \cite{gunnar} of the LSD for exchange and correlation energy.  The density used in equation 
\eqref{eq:exke0-2gap} and \eqref{T2*-2gap} is the one obtained from the same Kohn-Sham calculations as done for exact kinetic energy.
Also shown in Table \ref{tab-co-sh-sh} are the corresponding energies calculated from the 
ground state kinetic energy functional $T^{(0)}$ and its gradient correction $T^{(2)}$. 
As is evident from the table, the ground state based functionals underestimate the exact kinetic energies by a large amount
whereas the proposed functionals reduce the error significantly. This points to the correctness of  physics invoked to construct the functionals.

  Similarly, the functional can also be applied to the shell-shell structure  where the lowest energy orbitals are empty,
and the next orbitals are occupied followed by some empty orbitals and again orbitals are occupied. The corresponding excited state functional is 
obtained by substituting $\rho_1=0$ in equations \eqref{eq:exke0-2gap} and \eqref{T2*-2gap}. 
 This kind of excited states is quite interesting because the error of ground state kinetic energy functionals in these is very large. 
This is shown in Table \ref{tab-sh-sh}. On the other hand   excited state functionals of equations \eqref{eq:exke0-2gap} and 
\eqref{T2*-2gap} again give energies which are far superior as is evident from Table \ref{tab-sh-sh}. 

Finally we mention that we have also included the fourth
order correction in a manner similar to above. The results obtained with the second-order do not change much with the inclusion of the fourth order.

\section{Response function analysis}
 
  With this accurately constructed functionals  based on writing  the energy functional $E$ in terms of
 $k_1,k_2,k_3,\cdots $  i.e $E[k_1,k_2,k_3,\cdots]=E[k_1(\rho_1(r)),k_2(\rho_2(r)),k_3(\rho_3(r)),\cdots]$, 
we ask if it is necessary to write the energy functional in terms of $\rho_1, \rho_2 , \rho_3 $ or 
can it be written in terms of the density
 $\rho=(\rho_1  -\rho_2  + \rho_3 )$ directly.
Simple answer is that for homogeneous electron gas this cannot be done because the density for the ground and any excited-state of 
the system is the same. Thus information about the excited state cannot come from density alone as also suggested by the bi-functional nature \cite{levynag} of
excited state functionals. In this paper, we demonstrate more rigorously through response function analysis that excited-state energy functionals cannot be written in terms of
$\rho(\textbf{r})$ alone. We show that if we try to write the GEA for kinetic energy using total $\rho=(\rho_1  -\rho_2  + \rho_3 )$,  
it leads to kinetic energy density that diverges. We conclude that although it may be possible to write excited energy functional 
in terms of the corresponding density for inhomogeneous systems in  some other ways, it is most easily done if we instead employ the densities  $\rho_1, \rho_2, \rho_3$  as done above.

In the following, we first obtain the response function of non-interacting homogeneous electron gas and expand it to construct the kinetic energy functional correct up to the second-order
in $\vec{\nabla}\rho({\mathbf{r}})$. We then show that the correction grows exponentially in the asymptotic regions of a finite system.

 Suppose one could write the kinetic energy for an excited state of a homogeneous electron gas as $T^{*(0)}[\rho]$. The system is now made 
slightly inhomogeneous through a perturbation $V_{ext}(\mathbf{r})$.  The corresponding  kinetic energy up to the second order is
obtained by expanding kinetic energy in density changes and is given as  \cite{hk,kvasis,rmp},

 \begin{equation}
     T[\rho(\mathbf{r})] = T^{*(0)}[\rho] +
\frac{1}{2} \iint K(|\mathbf{r}-\mathbf{r}'|) \Delta \rho(\mathbf{r}) \ \Delta \rho(\mathbf{r}') d\mathbf{r} \  d\mathbf{r}'
 \label{Tnfnal}
 \end{equation}
 In the Fourier space, the total energy of the system,
using $\rho(\mathbf{q})  =  \int \Delta \rho(\mathbf{r}) e^{-i \mathbf{q} \cdot \mathbf{r}} d\mathbf{r}$, is 
\begin{eqnarray}
  E[\rho] = T^{*(0)}[\rho] + \frac{1}{2} \frac{1}{(2\pi)^{3}} \int K(\mathbf{q}) |\rho(\mathbf{q})|^2  d\mathbf{q} 
                    + \frac{1}{(2\pi)^{3}} \int  V_{ext}(\mathbf{q}) \rho(\mathbf{q}) d\mathbf{q}
  \label{Enfnal}
\end{eqnarray}
where $K(\mathbf{q})$ is the Fourier transform of $K(|\mathbf{r}-\mathbf{r}'|)$.
 Making the energy stationary with respect to $\rho(\mathbf{q})$ gives
$\rho(\mathbf{q}) = V_{ext}(\mathbf{q}) K(\mathbf{q})^{-1}$. $K(\mathbf{q})$ is obtained by identifying  $- K(\mathbf{q})^{-1}$ with $\chi^{(0)}(\mathbf{q})$,  where $\chi^{(0)}(\mathbf{q})$
is the response function of non-interacting homogeneous electron gas obtained using perturbation theory and is given as
 \begin{eqnarray}
      \chi(\mathbf{q})  =  \frac{1}{V}  \sum_{\vec{k} : occ}  \left[ \frac{1}{\varepsilon^0_{\vec{k}} - \varepsilon^0_{\vec{k}+\vec{q}}} 
+ \frac{1}{\varepsilon_{\vec{k}}^0 - \varepsilon^0_{\vec{k}-\vec{q}}} \right] + c.c 
\nonumber
%  \label{chiq-sum}
 \end{eqnarray}
where $\varepsilon^{(0)}_{\vec{k}} = \frac{k^2}{2}$ and $V$ is the volume over which periodic boundary conditions are applied.
 The response function above for excited state corresponding to Figure \ref{figure} is given in a simple way as 
 \begin{eqnarray}
   \chi^{(0)}_G(k_1,k_2,k_3;\mathbf{q}) & = &  \chi^{(0)}_G(k_1;\mathbf{q})  - \chi^{(0)}_G(k_2;\mathbf{q}) + \chi^{(0)}_G(k_3;\mathbf{q})
 \end{eqnarray}
 where $\chi^0_G(k;\mathbf{q}) $ is the response function for the ground state with Fermi wave vector $k$ and is given as
 \begin{equation}
   \chi^0_G(k;\mathbf{q}) =   - \frac{1}{\pi^2 q}  \biggr[ \frac{qk}{2} +
\left(   \frac{k^2}{2} - \frac{q^2}{8} \right)ln\left| \frac{q+2k}{q-2k} \right|   \biggl]
 \label{chi-gr}
 \end{equation}
Expanding $\chi^{(0)}_G(k_1,k_2,k_3;\mathbf{q})$ to order $q^2$ leads to the second order correction for kinetic energy. For excited states, small $q$-limit 
 gives
 \begin{equation}
    \chi^{(0)}(k_1,k_2,k_3;\mathbf{q}) \approx  - \frac{1}{\pi^2}  \left\lbrace k_1-k_2+k_3 -\frac{q^2}{12} \left[ \frac{1}{k_1} - \frac{1}{k_2} + \frac{1}{k_3} \right]
      \right\rbrace
    \nonumber
 \end{equation}
leading to 
 \begin{equation}
   K(\mathbf{q}) \equiv  - \chi^{(0)}(\mathbf{q})^{-1} =   \frac{\pi^2 }{(k_1-k_2+k_3)}
\left\lbrace 1+\frac{q^2}{12} \frac{1}{(k_1-k_2+k_3)}
 \left[ \frac{1}{k_1} - \frac{1}{k_2} + \frac{1}{k_3} \right] \right\rbrace
    \nonumber
 \end{equation}
The excited state kinetic energy functional is then obtained by  substituting $K(\mathbf{q})$ in \eqref{Tnfnal}, so that
  \begin{eqnarray}
      T[\rho(\mathbf{r})] & =  & T_0[\rho] +   \frac{\pi^2}{2}  \int  \frac{ \left|\Delta \rho(\mathbf{r})\right|^2}{(k_1-k_2+k_3)} d\mathbf{r}
+ \frac{\pi^2}{24} \int  \frac{\left|\nabla \rho(\mathbf{r})\right|^2}{(k_1-k_2+k_3)^2}
 \left[ \frac{1}{k_1} -\frac{1}{k_2} + \frac{1}{k_3} \right] d\mathbf{r} \ \ 
 \label{final1}
  \end{eqnarray} %$\left|\Delta \rho(\mathbf{r})\right|^2 $
 The second term in the expression above is identified as  $ \frac{ \delta^2 T^{*(0)}}{\delta\rho(\mathbf{r})\delta\rho{(\mathbf{r}')}} $ of functional $T^{*(0)}$
of equation \eqref{eq:exke0}. It is clear from the above that it cannot be written in terms of $\rho = \rho_1 - \rho_2 + \rho_3$ alone. 
To check its correctness if we let $k_1 = k_2 \ \textrm{and} \  k_3 = k_F$, this term goes over to $ \frac{ \delta^2 T_s^{(0)}}{\delta\rho(\mathbf{r})\delta\rho{(\mathbf{r}')}} $ 
of functional $T_s^{(0)}$ of equation \eqref{eq:tf}
for the ground state. The third term above gives the gradient correction to the kinetic energy up to the second order in $\vec{\nabla}\rho $. 
Upon substituting $k_1 = k_2 \ \textrm{and} \  k_3 = k_F$ this also goes correctly over to $\frac{1}{72} \int \frac{|\nabla\rho(\mathbf{r})|^2}{\rho(\mathbf{r})} d\mathbf{r}$, 
the gradient correction for the ground state. However, for excited states, this term is not well behaved for $r \rightarrow \infty$. 
In this limit, similar to ground state \cite{morrell,levpersah,katriel} densities, excited-state density also varies \cite{shamimCPL} as $ \rho(\mathbf{r}) 
\sim exp \left[ -2 \sqrt{-2 \ \varepsilon_{max}} \   r \right] $  where $\varepsilon_{max}$ is the highest occupied orbital energy. 
Hence $ k_3 \sim   exp \left( -\frac{2}{3} \left(- 2 \varepsilon_{max}\right)^{1/2} r \right)  $.
 For excited states   $ k_1 < k_2 < k_3 \  $ so  
 the gradient term in asymptotic limit is proportional to
%$\frac{ \left|\Delta \rho_1(\mathbf{r})\right|^2}{k_3} $ and 
$\frac{ \left|\nabla \rho_3(\mathbf{r})\right|^2}{k_1 k_3^2} \sim \frac{k^4_3}{k_1}$.
In terms of orbital energies  $ \varepsilon_1, \varepsilon_2, \varepsilon_3 $ %of the upper most orbitals of the three regions 
corresponding to $k_1, k_2$ and $k_3$, shown in Figure \ref{figure}, this is given by  $exp \left[ \frac{2}{3} \left( \sqrt{-2 \varepsilon_1} -4 \sqrt{-2 \varepsilon_3 } \right) r \right]$.
Since,  $ |\varepsilon_1| > |\varepsilon_2| > |\varepsilon_3| \  $, the gradient term diverges in the asymptotic limit for $|\varepsilon_1| > 16|\varepsilon_3|$
if we  insist on expanding the kinetic energy functional in terms of the density. This diverging behavior of the kinetic energy density is also an explicit 
demonstration of the lack of Hohenberg-Kohn theorem for excited states \cite{gaudoin}. Thus additional input is needed \cite{samalCPL} 
if one wishes to employ excited state density for calculation. As shown by our results in the beginning of the paper, this is done easily by dealing
 with $\rho_1, \rho_2 $ and $\rho_3$ separately right from the beginning. Indeed, the exchange energy functional constructed in a similar manner also leads to 
accurate results for excited state energies \cite{samal2,hsamal,samalh,aip}.

\section{Concluding Remarks : }
 Exchange and kinetic energy LDA functionals constructed by splitting $k$-space for homogeneous electron gas have been shown to be accurate in the past.
Analysis of the kinetic energy functional based on the response function of an excited-state of HEG suggests that it is not possible to construct LDA functionals that
are dependent on excited state density alone. Splitting the $k$-space and working with the corresponding core and shell densities separately thus provides a method of constructing 
density functionals for excited states and may pave the way to ground-state like density functional calculations for excited-states too.
 
{\bf Acknowledgment:} We thank Md. Shamim for useful discussions.

\newpage

\begin{table}
\caption{\label{tab-co-sh-sh} Approximate kinetic energies (in atomic units) obtained for excited states of some atoms 
through the use of ground state functionals $T^{(0)}$ and  $T^{(0)}+T^{(2)}$ (equations~\eqref{eq:tf} and~\eqref{eq:gea2}) 
and those obtained from $T^{*(0)}$ and  $T^{*(0)}$+$T^{*(2)}$ (equations~\eqref{eq:exke0-2gap} and~\eqref{T2*-2gap} ).
Comparison is made with the exact kinetic energy resulting from the solution of Kohn-Sham equation for excited states.
}
\vspace{0.2in}
\begin{tabular}{p{4.20cm}p{2.0cm}p{2.0cm}p{2.0cm}p{2.0cm}p{2cm}}
\hline
\multirow{2}{*}{Atom}&\multicolumn{2}{c}{ZEROTH ORDER}&\multicolumn{1}{c}{}&\multicolumn{2}{c}{SECOND ORDER} \\
\cline{2-3}
\cline{5-6}
& $T^{(0)}$  &  $T^{*(0)}$  &  $T^{(Exact)}$  &    $T^{*(0)}$+$T^{*(2)}$ &  $T^{(0)}+T^{(2)}$ \\
\hline 
$Na(1s^{2}2s^{0}2p^{4}3s^{0}3p^{5})$ &131.024&138.916&150.803&152.797&142.480\\
%&(13.11)&(7.88)&&(1.32)&(5.52)\\
$Na(1s^{2}2s^{0}2p^{5}3s^{0}3p^{4})$ &133.511&142.007&153.377&155.897&144.881\\
%&(12.95)&(7.41)&&(1.64)&(5.54)\\
$Mg(1s^{2}2s^{0}2p^{6}3s^{0}3p^{4})$ &168.132&179.525&191.196&196.374&181.698\\
%&(12.06)&(6.10)&&(2.71)&(4.97)\\
$Al(1s^{2}2s^{0}2p^{6}3s^{0}3p^{5})$ &203.234&217.535&231.138&237.663&219.206\\
%&(12.07)&(5.88)&&(2.82)&(5.16)\\
$Si(1s^{2}2s^{0}2p^{6}3s^{0}3p^{6})$ &242.410&260.000&275.573&283.740&260.998\\
%&(12.03)&(5.65)&&(2.96)&(5.29)\\
\hline
\end{tabular}
\end{table}

\begin{table}
\caption{\label{tab-sh-sh} Approximate kinetic energies (in atomic units) obtained for pure excited states of some atoms 
through the use of ground state functionals $T^{(0)}$ and  $T^{(0)}+T^{(2)}$ (equations~\eqref{eq:tf} and~\eqref{eq:gea2}) 
and those obtained from $T^{*(0)}$ and  $T^{*(0)}$+$T^{*(2)}$ (equations~\eqref{eq:exke0-2gap} and~\eqref{T2*-2gap} ).
Comparison is made with the exact kinetic energy resulting from the solution of Kohn-Sham equation for excited states.
}
\vspace{0.2in}
\begin{tabular}{p{4.20cm}p{2.10cm}p{2.10cm}p{2.10cm}p{2.10cm}p{2.1cm}}
\hline
\multirow{2}{*}{Atom}&\multicolumn{2}{c}{ZEROTH ORDER}&\multicolumn{1}{c}{}&\multicolumn{2}{c}{SECOND ORDER} \\
\cline{2-3}
\cline{5-6}
& $T^{(0)}$  &  $T^{*(0)}$  &  $T^{(Exact)}$  &    $T^{*(0)}$+$T^{*(2)}$ &  $T^{(0)}+T^{(2)}$ \\
\hline
$B(1s^{0}2s^{0}2p^{3}3s^{0}3p^{2})$ &2.794&6.308&7.308&6.571&3.025\\ 
%&(61.77)&(13.69)&&(10.09)&(58.61)\\
% $N(2s^{2}2p^{3}3s^{0}3p^{2})$
$N(1s^{0}2s^{2}2p^{3}3s^{0}3p^{2})$ &9.593&17.393&20.962&18.201&10.345\\
%&(54.24)&(17.02)&&(13.17)&(50.65)\\
% $F(2p^{4}3s^{0}3p^{5})$
$F(1s^{0}2s^{0}2p^{4}3s^{0}3p^{5})$ &18.676&34.908&36.703&36.316&19.788\\ 
%&(49.12)&(4.89)&&(1.05)&(46.09)\\
% $Ne(2s^{2}2p^{4}3s^{0}3p^{4})$
$Ne(1s^{0}2s^{2}2p^{4}3s^{0}3p^{4})$ &31.164&50.416&55.958&52.514&32.991\\
%&(44.31)&(9.90)&&(6.15)&(41.04)\\
% $Mg(2s^{2}2p^{4}3s^{0}3p^{6})$
$Mg(1s^{0}2s^{2}2p^{4}3s^{0}3p^{6})$  &50.125&80.608&89.788&84.062&52.894\\
%&(44.17)&(10.22)&&(6.38)&(41.09)\\
% $Mg(2s^{2}2p^{6}3s^{0}3p^{4})$
$Mg(1s^{0}2s^{2}2p^{6}3s^{0}3p^{4})$ &61.526&92.917&97.778&96.224&64.538\\
%&(37.08)&(4.97)&&(1.59)&(34.00)\\
% $Si(2s^{2}2p^{6}3s^{0}3p^{6})$
$Si(1s^{0}2s^{2}2p^{6}3s^{0}3p^{6})$ &93.290&139.872&146.527&144.983&97.636\\
%&(36.33)&(4.54)&&(1.05)&(33.37)\\
\hline
\end{tabular}
\end{table}

\end{document}